\documentclass{aa}  

\usepackage{graphicx}
\usepackage{txfonts}
\usepackage[colorlinks=true,urlcolor=magenta]{hyperref}
\begin{document}

   \title{Evolution of a Streamer-Blowout CME as Observed by Imagers on  Parker Solar Probe  and the Solar Terrestrial Relations Observatory}

\titlerunning{Streamer-blowout CME in PSP and STEREO}


   \date{Received 00000000; accepted 00000000, 2020}

 
\abstract 
{On 26-27 January 2020, the wide-field imager WISPR on Parker Solar Probe (PSP) observed a coronal mass ejection (CME) from a distance of approximately 30 R$_{\odot}$ as it passed through the instrument’s $95^\circ$ field-of-view, providing an unprecedented view of the flux rope morphology of the CME’s internal structure. The same CME was seen by STEREO, beginning on 25 January.}
{Our goal was to understand the origin and determine the trajectory of this CME.}
{We analyzed data from three well-placed spacecraft: Parker Solar Probe (PSP), Solar Terrestrial Relations 
Observatory-Ahead (STEREO-A), and Solar Dynamics Observatory (SDO). The CME trajectory was determined using the method described in \citet{Liewer2020} and verified using simultaneous images of the CME propagation from STEREO-A. The fortuitous alignment with STEREO-A also provided views of coronal activity leading up to the eruption. Observations from SDO, in conjunction with potential magnetic field models of the corona, were used to analyze the coronal magnetic evolution for the three days leading up to the flux rope ejection from the corona on 25 January.}
{We found that the 25 January CME is likely the end result of a slow magnetic flux rope eruption that began on 23 January and was observed by STEREO-A/Extreme Ultraviolet Imager (EUVI). Analysis of these observations suggest that the flux rope was apparently  constrained in the corona for more than a day before its final ejection on 25 January. STEREO-A/COR2 observations of swelling and brightening of the overlying streamer for several hours prior to eruption on January 25 led us to classify this as a streamer-blowout CME. The analysis of the SDO data suggests that restructuring of the coronal magnetic fields caused by an emerging active region led to the final ejection of the flux rope.}
{}


\author{ P. C. Liewer \inst{1}
\and J. Qiu  \inst{2}
\and A. Vourlidas \inst{3}
\and  J. R. Hall \inst{1} 
\and  P. Penteado\inst{1}}

\institute{Jet Propulsion Laboratory, California Institute of Technology, Pasadena, CA, 91109,
USA\\
\and
Montana State University, Bozeman, MT, 59717 ,USA\\
\and
The Johns Hopkins University Applied Physics Laboratory, Laurel, MD 20723, USA }

\maketitle
%

\begin{figure}
          \resizebox{\hsize}{!}{\includegraphics{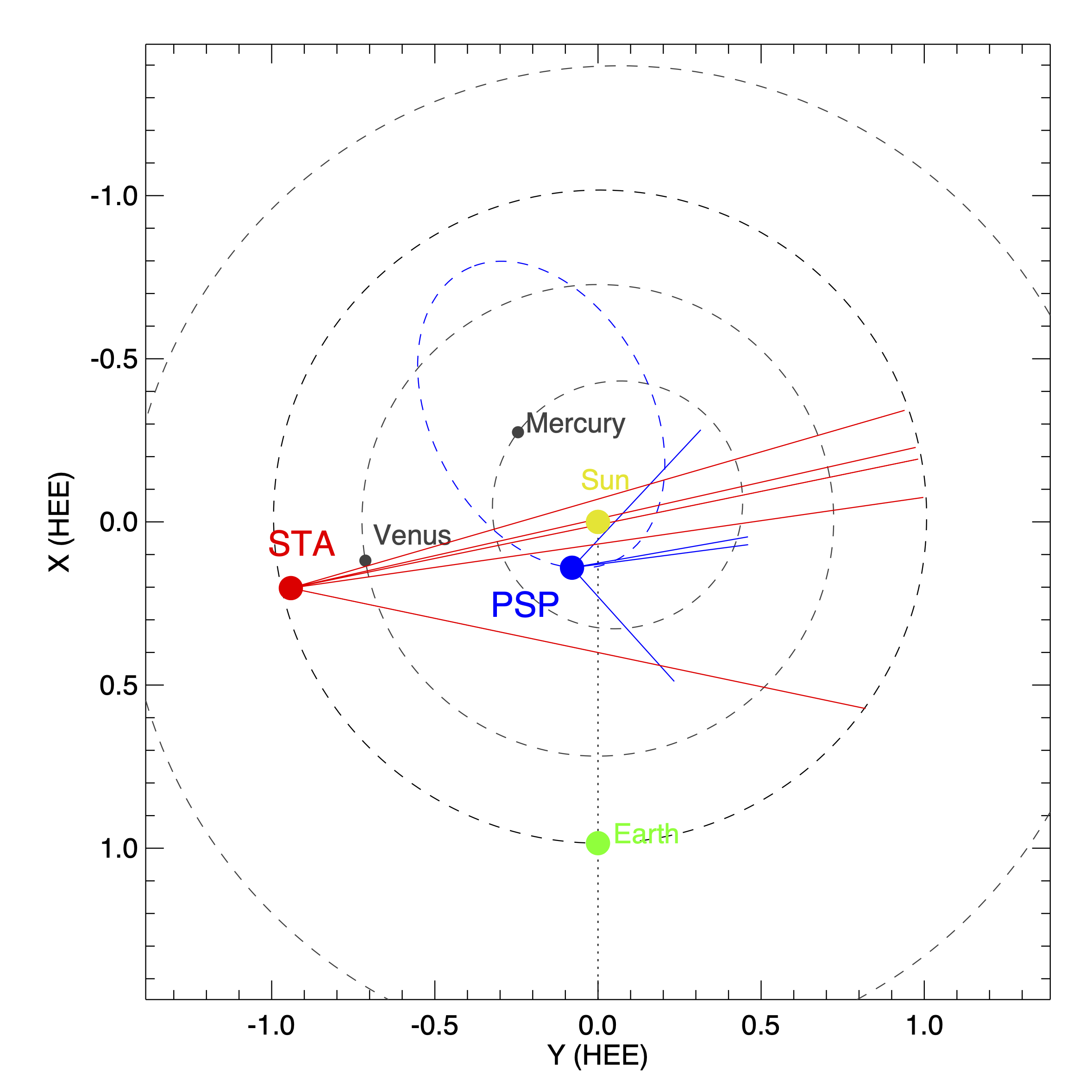}}
   \caption{ Polar plot showing the locations of the three spacecraft (PSP, STA and SDO(Earth)) on 27 January 2020 at 0 UT. 
   The fields-of-view of the WISPR-I and WISPR-O telescopes on PSP and COR2 and HI-1 on STA are indicated by solid lines. The plot is in the Heliocentric Earth Ecliptic (HEE)  coordinate frame and distances are in AU.}
  \label{fig:fig0}
\end{figure}

  \begin{figure*}
   \centering
   \includegraphics[width = 17 cm]{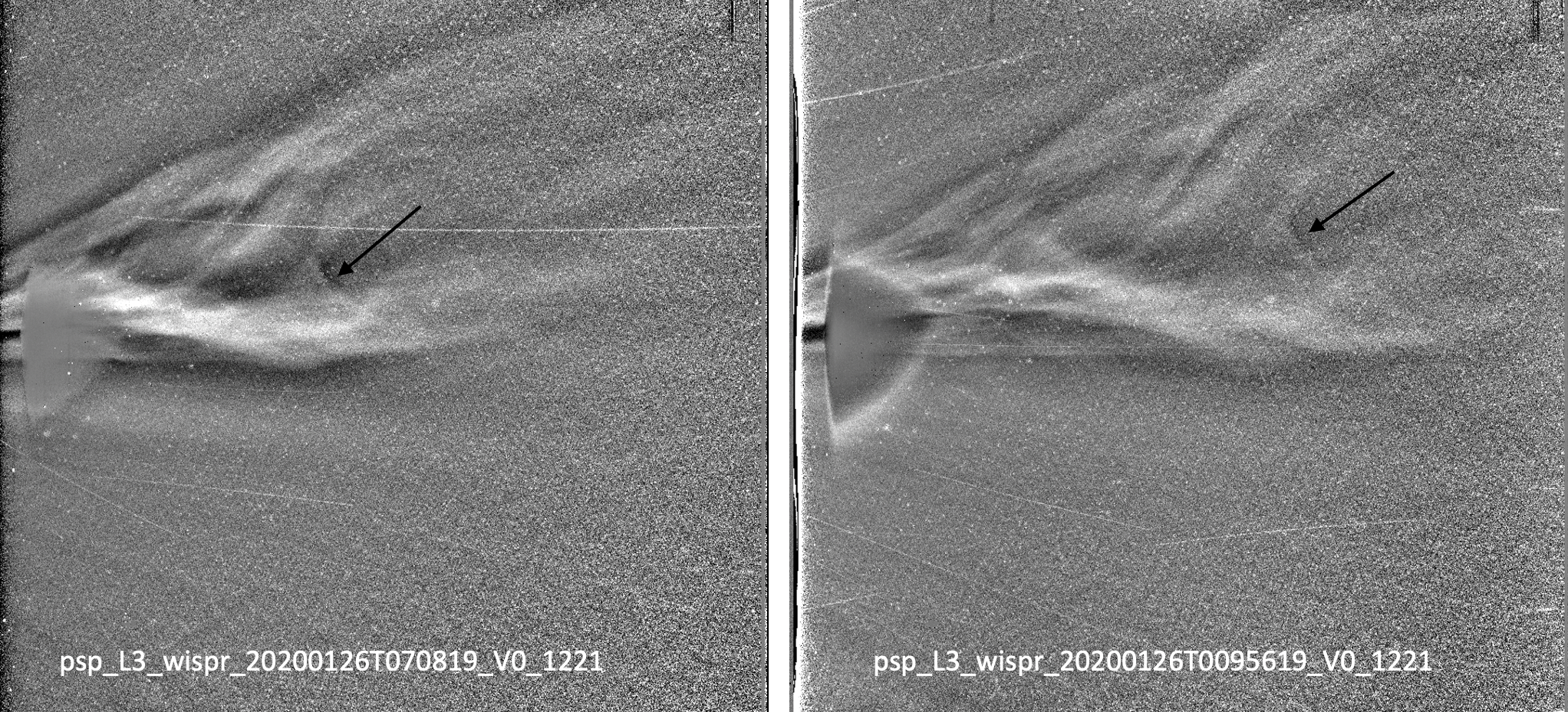}
   \caption{The CME at two times, 26 January 2020 at 7 UT (left) and 10 UT (right), as seen in WISPR Inner telescope.   The inner edge (left side) is $13.5^{\circ}$ from the Sun center. The cavity, as well as striations along the length of the flux rope, are evident.   The back of the cavity is indicated by black arrows. Denser material at the back of the flux rope is also evident. The images  are Level-3 data with the background removed (see text). The white streaks are caused by sunlight reflecting off debris created by dust striking the spacecraft. The triangular feature near the inner edge {\bf of } the images is an artifact resulting from pixel saturation. }
              \label{fig:fig1}
    \end{figure*}

\section{Introduction}

Streamer-blowout CMEs (SBO-CMEs) are typically slow CMEs which originate in the streamer belt, taking streamer material along as they exit the corona, and leaving an evacuated streamer behind. The streamer itself is generally seen to brighten and swell prior to the CME eruption, with considerable coronal outflow following in its wake 
\citep{Sheeley1982, Illing1986, Howard1986}. From a study of 909 LASCO SBO-CMEs, \citet{Vourlidas2018} found that SBO-CMEs are associated with polarity inversion lines away from active regions and are often accompanied by a prominence eruption. The observed position angle in coronagraph images is generally that of the streamer belt and heliospheric current sheet.
 
On 26-27 January 2020, the Wide-field Imager for Solar Probe \citep[WISPR;][]{Vourlidas2016} on Parker Solar Probe \citep[PSP;][]{Fox2016} 
observed a CME closely  as it passed through the telescope's 95$^{\circ}$ wide field-of-view. The WISPR instrument has two white-light 
telescopes, WISPR-I (inner; covering approximately $13.5^{\circ} - 53.0^{\circ}$ elongation from Sun center) 
and WISPR-O 
(outer; covering approximately $50.5^{\circ} - 108.5^{\circ}$ elongation from Sun center). The transverse 
extent is approximately $50^{\circ}$; the inner edge points a fixed $13.5^{\circ}$ from Sun center.
 
The fortuitous alignment with the Solar Terrestrial Relations Observatory (STEREO) Ahead spacecraft \citep[STA;][]{Kaiser2008} at this time provided views of coronal activity leading up to the eruption, and, based on the STA/COR2 observations of swelling and brightening of the streamer for several hours prior to eruption, we classify this as a SBO-CME. PSP was approximately 30 solar radii from the Sun during the observation, providing an unprecedented view of the flux rope morphology of the CME’s internal structure. We used the WISPR data to determine the trajectory of the CME by means of the tracking and fitting technique introduced in \citet{Liewer2020}. The CME eruption itself was observed by the STA/Sun Earth Connection Coronal and Heliospheric Investigation \citep[SECCHI; ][]{Howard2008} coronagraphs as a near-limb CME on 25 January and the propagation was seen in both COR2 and HI-1 images. The STA’s second view-point was also used to verify the trajectory determined from the WISPR data using a novel triangulation technique.
 
The CME ejection seemed to be associated with the newly emerged active region, National Oceanic and Atmospheric Administration (NOAA) active region AR 12757. However, data from the Extreme Ultraviolet Imager \citep[EUVI;][]{wuelser2004} on-board STA suggest that the flux rope may have been sitting high in the corona for more than a day before the ejection. AR 12757 may have altered the coronal magnetic structure in a manner that allowed the existing flux rope to escape.  To investigate this scenario, we analyzed data from the Solar Dynamics Observatory \citep[SDO;][]{Pesnell2012}, which, during this period, was separated by about 80$^{\circ}$ from STA, and provided disk observations of the solar activity underlying the CME source region. These observations were compared with potential-field source-surface (PFSS) models of the corona to help understand the source region and the sequence of events. The trajectory determined from the WISPR data suggests that the source region was located significantly west of AR~12757.
To provide context for the analysis presented below, the positions of the three spacecraft whose data {\bf were} used -- PSP, STA and SDO (at Earth)-- are shown in the polar plot of 
Fig.~\ref{fig:fig0}.  The fields-of-view of the WISPR-I and WISPR-O telescopes on PSP and COR2 and HI-1 on STA are indicated by solid lines emanating from their spacecraft. 
These are the positions on 27 January at 0 UT. The dashed blue line shows PSP's orbit for this time period. 

The organization of the paper is as follows: In Section 2, we present the WISPR observations of the CME on 25-26 January, 2020 and the trajectory of the CME determined from these data. The observation of the CME eruption and propagation by STA/COR2 and HI-1 are also presented in Section 2. In Section 3, we present observations and analysis chronicling the formation and partial eruption and final ejection of the CME, the analysis of the corresponding SDO observations, and the comparison with the PFSS models. In Section 4 we present the results for determination of the CME location using triangulation between STA and WISPR and compare these with the trajectory of the CME determined from the WISPR data alone. Section 5 contains a summary and discussion of the results.

\section{White Light Observations of the CME Propagation}

\subsection{WISPR Observations of the CME 26-27 January and Determination of the Trajectory}

The CME, as seen in the WISPR-I telescope, is shown at two times on 26 January 2020 in Fig.~\ref{fig:fig1}. These are Level-3 images that have been calibrated in units of Mean Solar Brightness (MSB), with bias, stray light and vignetting corrections applied, and a background subtracted. (The publicly released data, and the descriptions of the data products, can be found at https://wispr.nrl.navy.mil/wisprdata.)  The white streaks in the WISPR images, here and the images below, are caused by sunlight reflecting off debris created by dust striking the spacecraft.  The images show several features of the CME flux rope morphology.  The cavity is visible, but no bright leading edge or shock is seen, presumably because of the low speed of the CME \citep[e.g.][]{Vourlidas2013}. The fact that the cavity is visible suggests that the axis of the CME is more or less perpendicular to  the image plane, and the rib-like striations surrounding the cavity suggest that the flux rope is at a slight yaw to this plane, allowing us to see structure along the flux rope axis. The cavity is followed by considerable bright dense material that presumably lies at the bottom of the flux rope cavity and is distributed along the axis.  At this time, the CME must be near
 the location in the sky where the Thomson-scattered signal into the telescope is maximum to allow such details to be seen;  the region of maximum scattering, along a given line of sight, is an {\bf extended region around the position} where the ray intersects the surface a sphere, called the Thomson sphere \citep{Vourlidas2006, Vourlidas2016}, which has a diameter extending from the Sun to WISPR.

\begin{figure*}
 
   \centering
   \includegraphics[width = 12 cm]{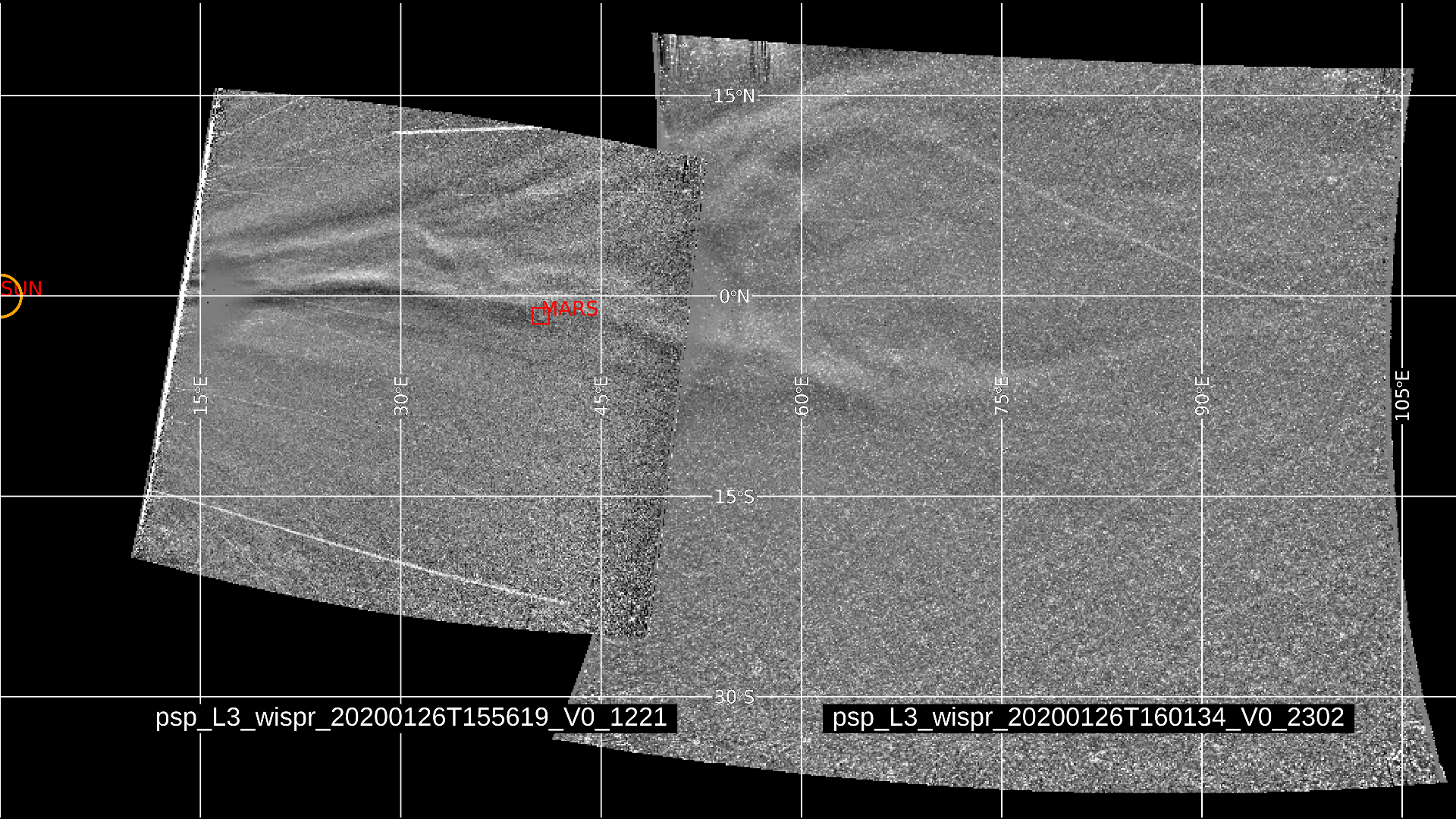}
   \includegraphics[width = 12 cm]{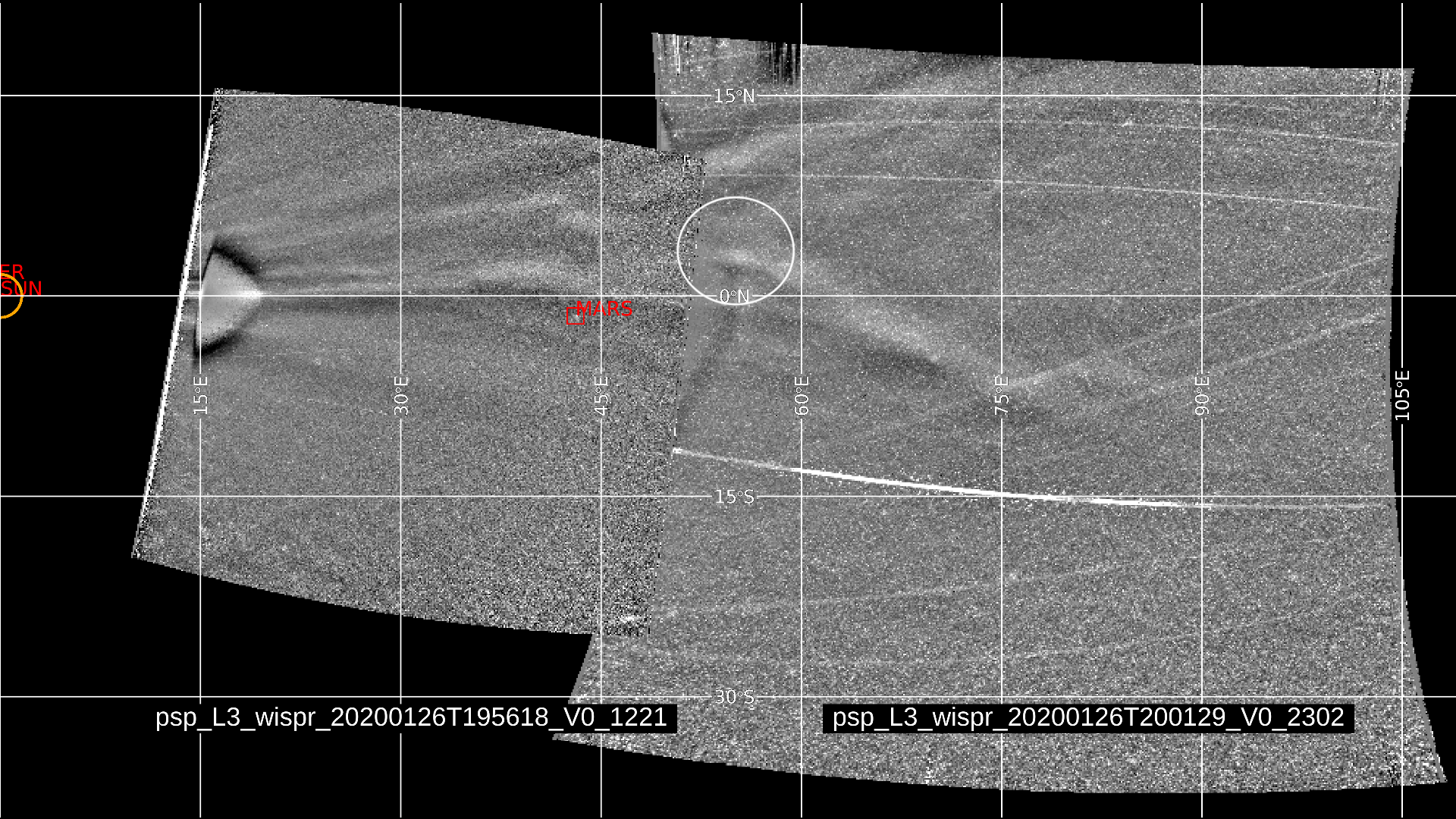}
   \caption{The CME at two times (26 January at 16 and 20 UT) seen in a projection combining the inner and outer telescope images. Visible in the top panel, above the CME cavity, is streamer material that has been pushed up by the CME.  In the lower panel, the re-formed streamer behind the CME is visible, and a diffuse V-shape feature is  visible behind the cavity that may mark the reconnection site  (white circle). 
The position of Mars is labeled in red. 
   The coordinate system used here is the observer-centric PSP orbital frame,  described in the text. The PSP orbit plane is labeled $0^{\circ}$N and the Sun (yellow circle, shown to scale) is located at (0,0). 
   A movie of the CME in this projection is available \href{http://sd-www.jhuapl.edu/secchi/paper_repo/psp_L3_wispr_join_SPP_ORBIT_grid_15fps.mp4}{online}. }
              \label{fig:fig2}
\end{figure*}

Fig.~\ref{fig:fig2} shows the CME at a later time on 26 January as it extends across the fields-of-view of both telescopes. 
In this figure, the two images have been reprojected into a coordinate system that we call the PSP orbital coordinate system \citep[see ][]{Liewer2020}. 
This observer-centric projective cartesian system is defined by two vectors: the Sun-spacecraft vector and the PSP velocity vector; together they define the PSP orbit plane. One angle is measured out of the PSP orbit plane and the other angle parallel to the plane; the angles are referenced to the midpoint of the inner telescope FOV. This frame is similar to  the HelioProjective Cartesian (HPC) coordinate system in that both are cartesian, observer-centric and projective. The HPC system uses the Sun\,--\,spacecraft vector and a vector perpendicular to this in the plane containing both the Sun-spacecraft vector and the solar north axis \citep{Thompson2006}. In both systems, the coordinates of the solar north vector vary. In this image and the movie, the Sun is at (0,0).  In this system density structures that might pass over the PSP spacecraft can be seen traveling in the PSP orbit plane.

In both images of Fig.~\ref{fig:fig2}, the CME cavity and dense material are still visible and, in addition, streamer material, which has been pushed aside, is visible above the CME structure.  In the lower image, the reformed streamer can be seen behind the CME, and a diffuse V-shaped feature is visible at the back of the cavity  (white circle) that may represent the reconnection site behind the CME. A movie of the CME in PSP orbital frame covering 26-27 January, 2020 is available \href{http://sd-www.jhuapl.edu/secchi/paper_repo/psp_L3_wispr_join_SPP_ORBIT_grid_15fps.mp4}{online}.

Using the WISPR images, we determined the trajectory of the CME using the method described in \cite{Liewer2020}. Briefly, a CME “feature” is tracked in a time sequence of images and the track is fit to analytic expressions relating the feature’s position in the image as a function of time to its trajectory in the Heliocentric Inertial (HCI) coordinate system. The analytic expressions are derived assuming the feature moves radially at a constant velocity. 
 The HCI coordinate system  is an inertial system, centered on the Sun, with the $z$-axis aligned with  solar north axis; latitude is measured from the $z$-axis and longitude is measured in the the solar equatorial plane \citep{Thompson2006}. In this system, the two trajectory angles remain constant for radial motion. 
The four trajectory parameters determined using this technique are the velocity, the two angles, and the radial distance from the Sun at the time tracking begins.  This technique is related to methods used to determine feature’s trajectories using J-maps \citep{Sheeley1999, Sheeley2008, Rouillard2008, Conlon2014}, but extended to include the effects on the WISPR images caused by PSP’s rapid, highly elliptical orbit \citep{Liewer2019, Liewer2020}.
 
 \begin{figure*}
   \includegraphics[width = 17 cm]{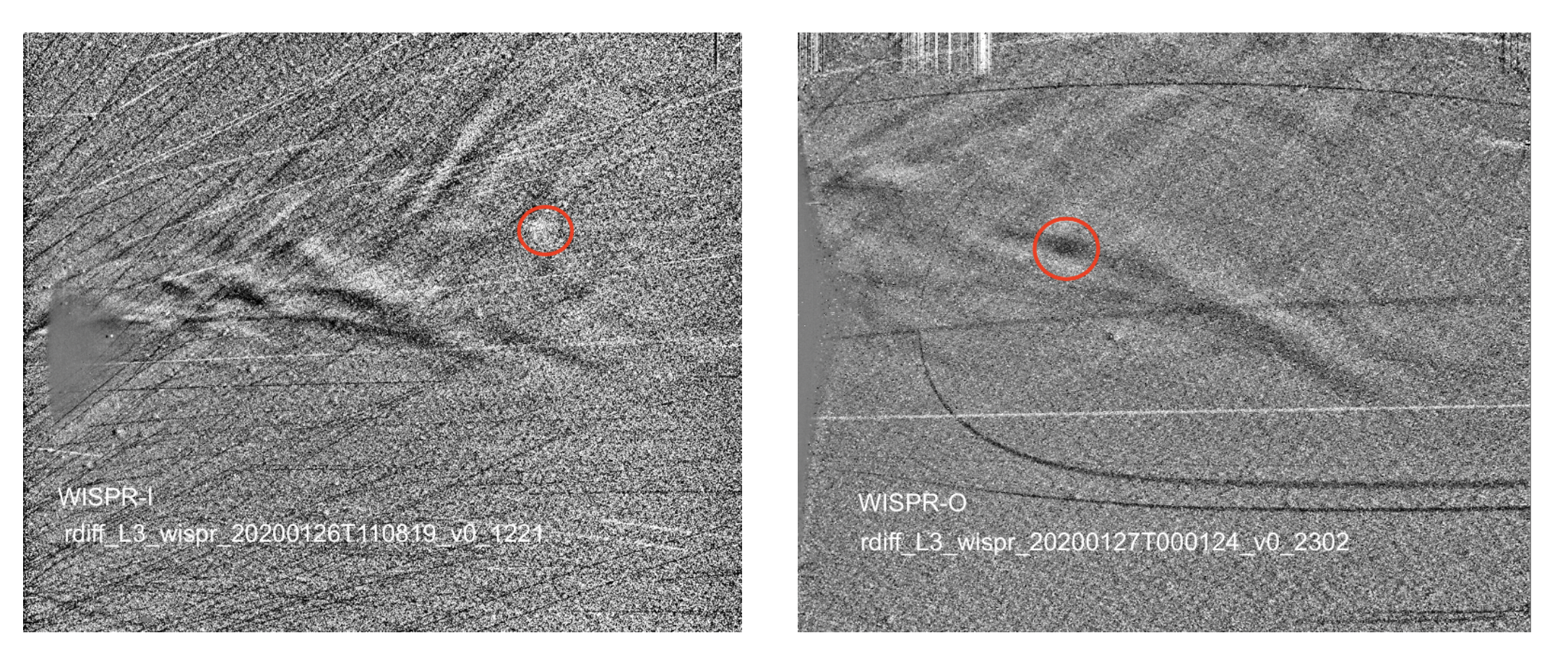}
   \caption{Two distinct features that were tracked in a series of images to determine the trajectory of the CME using the technique of \cite{Liewer2020}; both images are Level-3 running-differenced images. Left: A WISPR-I image on 26 January 11 UT with the tracked feature (bright V at back of the cavity) circled in red. Right: A WISPR-O differenced image on 27 January 0 UT  with the feature (dark spot below bright V-shape) circled. The resulting trajectories are summarized in Table 1. }
              \label{fig:fig3}
\end{figure*}

\begin{table*}
\caption{Trajectory Results from Tracking and Fitting}
\label{table:1}
\centering
\begin{tabular}{l c c c c}     
\hline\hline
Feature Tracked & \multicolumn{2}{c}{Bright V at back of Cavity WISPR-I} & \multicolumn{2}{c}{Dark Spot by V WISPR-O}\\
\hline 
Parameter & mean & uncertainty & mean & uncertainty \\
\hline
tracking start time $t_0$ & \multicolumn{2}{l}{2020-01-26  T03:33} & \multicolumn{2}{l}{2020-01-26 T20:04}\\
distance from Sun (R$_{\odot}$)	& 14  &	0.1 & 30.3 & 0.3 \\
velocity (km s$^{-1}$)          & 278 & 6	& 248  & 16 \\
HCI longitude (degree)&	57	& 2	 &	65 &	2 \\
HCI latitude (degree) &	1	& 2	&	2 &	2 \\
HEE longitude (degrees) at 2020-01-26T00:00	& 7.4 & & 15.7 & \\
HEE latitude (degrees) at 2020-01-26T00:00	& 7 &	 &	9 & \\
\hline 
\end{tabular}
\end{table*}

Fig.~\ref{fig:fig3} shows two features that were tracked, one in WISPR-I and one in WISPR-O, and their trajectories determined. No feature was found that could be accurately tracked in the FOVs of both telescopes.
These images, and the sequence used in the tracking, are Level-3 running difference images. We use running-difference images because it makes it possible to track features over a longer period of time. The circled feature in the WISPR-I image on the left corresponds to a bright feature at the back of the cavity seen in the upper panel of Fig.~\ref{fig:fig1}. The WISPR-I feature, the bright V at the back of the cavity, was tracked in 12 images covering 10.5 hours. The results of the tracking and fitting for this feature gave a velocity of 278$ \pm $6  km/s and HCI longitude and latitude of 57$^{\circ} \pm  2^{\circ}$ and $1^{\circ} \pm 2^{\circ}$, respectively. 
Note that the HCI angles are constant since radial motion has been assumed in obtaining this trajectory. These 
HCI angles correspond to Heliocentric Earth Ecliptic (HEE) angles of (7.4$^{\circ}$, 7.0$^{\circ}$) for 
the date 26 January 2020 at 12:00 UT.  The HEE system uses the Earth Ecliptic plane as the $x-y$ plane ($0^{\circ}$ latitude), with the $x$-axis ($0^{\circ}$ longitude) pointing to Earth.
Thus, the CME flank could have impacted Earth. The feature in 
the WISPR-O image on the right is a dark spot just below the bright V-shaped figure in the WISPR-O 
image on the lower panel of Fig.~\ref{fig:fig2}. This feature was tracked in 13 images covering 10 hours.The results of the tracking and fitting gave HCI longitude and latitude = 
(65$^{\circ} \pm $2$^{\circ}$, 2$^{\circ} \pm  2^{\circ}$) which corresponds to Heliocentric Earth Ecliptic 
(HEE) angles of (15.7$^{\circ}$, 9.0$^{\circ}$) for the date 26 January 2020 at 12:00 UT. These tracking 
and fitting results (velocity, angles, and radius at time tracking starts) are summarized in Table 1.  
Averaging the velocities and angles of these two solutions gives v=263$ \pm $20 km/s 
and HCI longitude and latitude = ($61^{\circ} \pm 5^{\circ}, 1^{\circ} \pm  2^{\circ}$).

\begin{figure}
          \resizebox{\hsize}{!}{\includegraphics{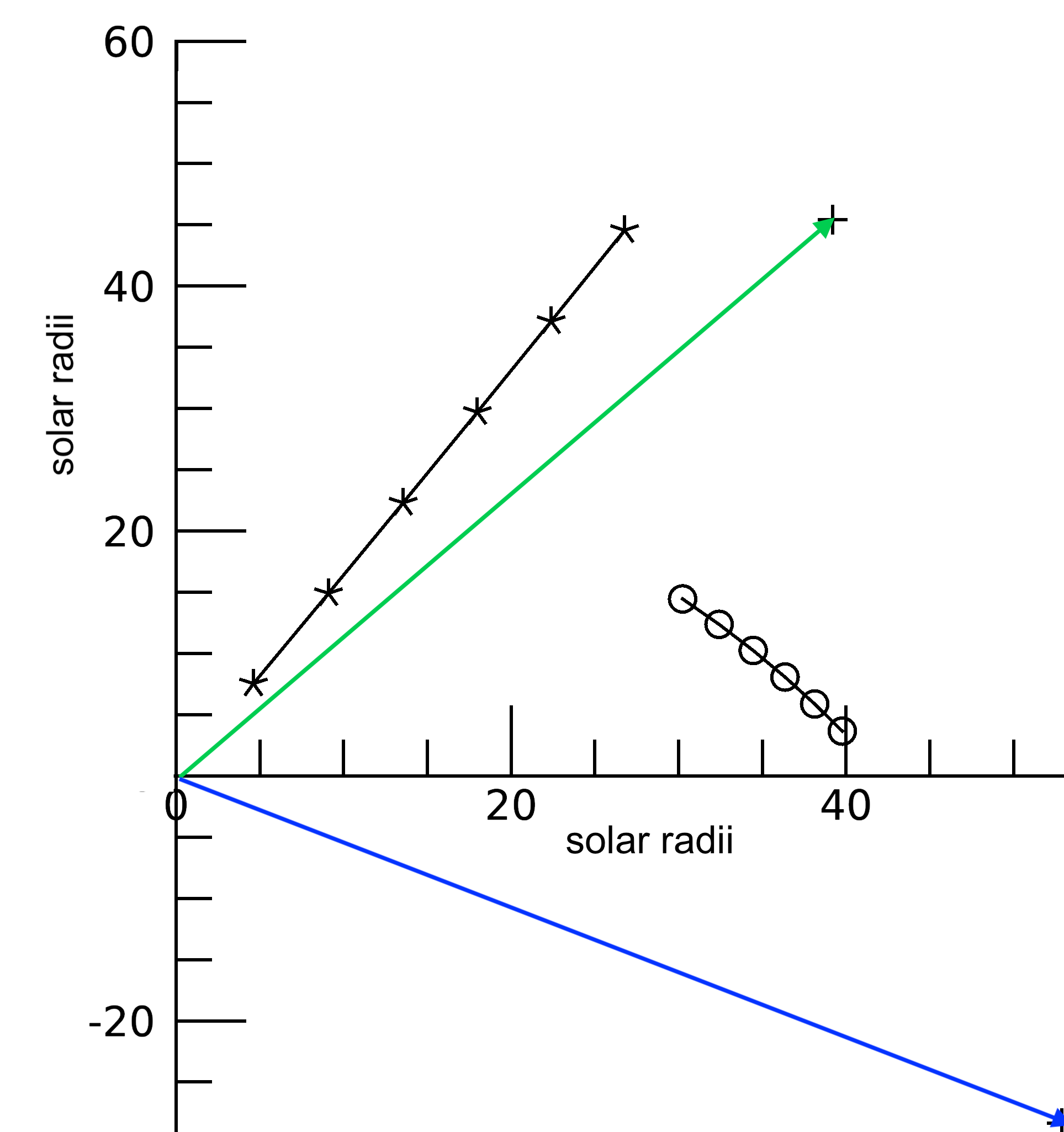}}
   \caption{ Polar plot showing the locations (HCI longitudes and distances from the Sun in solar radii) at several times for PSP (o) and the predicted locations of the CME (*) using the trajectory found from the feature tracked in the WISPR-I images. The locations of each are plotted at the same six times:  every 6 hours from 26 January  at 0 UT to 27 January at 6 UT, the approximate time interval that the CME was observed by WISPR.  PSP, heading toward perihelion, is approaching the path of the CME. The blue arrow shows the direction to STEREO-A and the green arrow the direction to Earth at the start of the time interval, 26 January  at 0 UT.}
              \label{fig:fig4}
\end{figure}

The polar plot in Fig.~\ref{fig:fig4} shows the predicted trajectory (Table 1) for the bright feature at the back of the cavity of the CME (*) and the trajectory of PSP(o) in the HCI coordinate frame with distances in solar radii and HCI longitude measured from the x-axis. Their locations are plotted every 6 hours from 26 January at 0 UT to 27 January at 6 UT, the approximate time interval the CME was observed by WISPR. From this, we see that, according to our predicted trajectory, PSP is approaching the CME, starting about 35 R$_{\odot}$ away and decreasing to about 30 R$_{\odot}$.  The CME is predicted to be very close to WISPR’s Thomson surface at the beginning, which explains the detailed observation of the internal structure, but it moves beyond the surface towards the end of the time interval. The trajectory solution predicts the separation between STA and the CME to be about 90$^{\circ}$ on 26 January  at 0 UT, consistent with the CME's appearance as a limb CME for STA. The arrows on the plot point in the direction to STA (blue, HCI longitude =$ -28^{\circ}$) and to Earth (green, HCI longitude = 49$^{\circ}$) at this same time. Using the velocities and distances from these trajectories to propagate the CME back to the Sun indicates the CME left the Sun before 17 UT on 25 January. How much before would depend on the CME's unknown acceleration profile.

\subsection{STEREO-A COR2 and HI-1 Observations of the CME}
 
\begin{figure*}
   \centering
    \includegraphics[width = 15 cm]{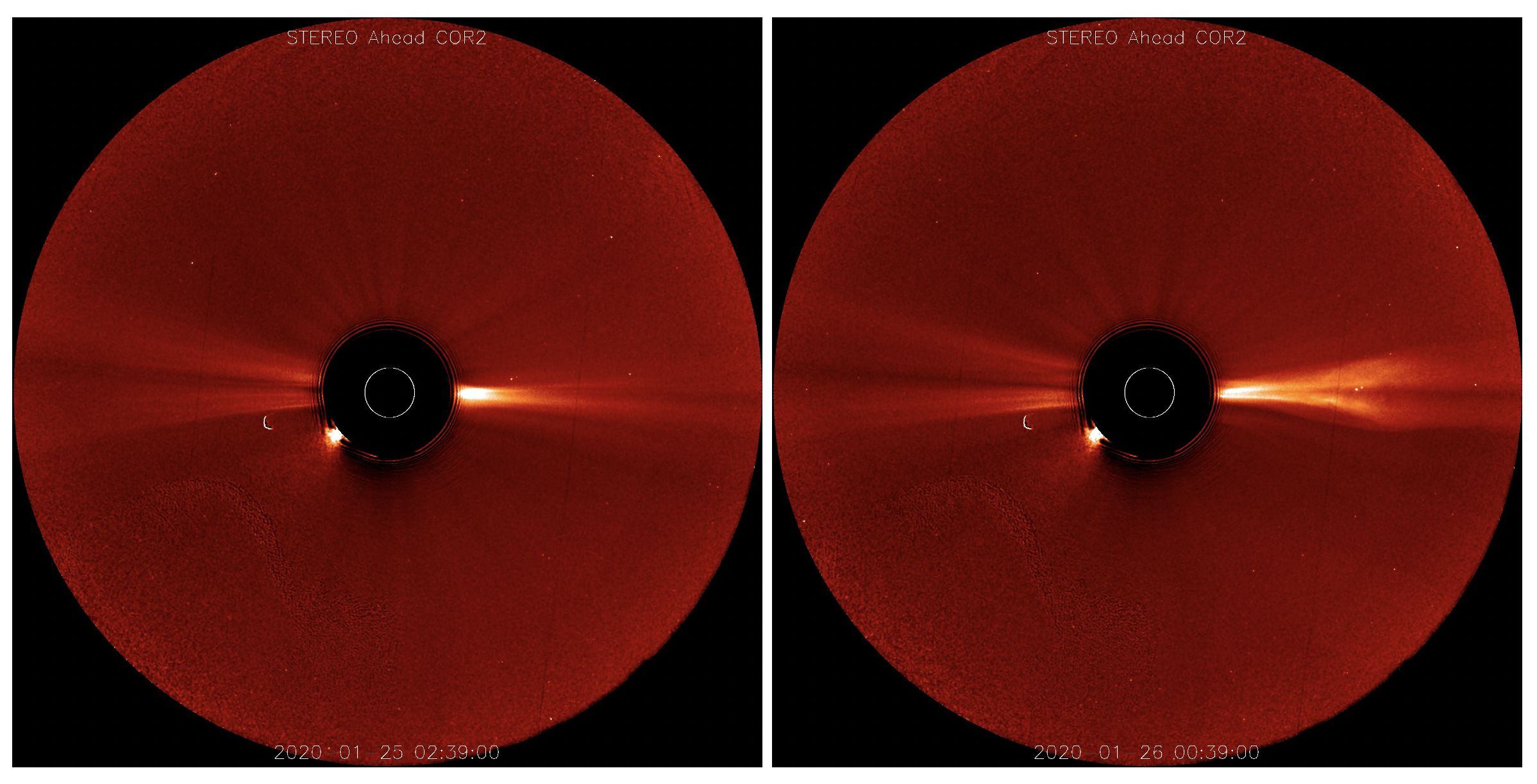}
   \caption{STEREO-A/COR2  images of the streamer before the eruption  on 25 January at 0:2:39~UT (left) and  of the streamer-blowout CME on 26 January at 00:36~UT (right) . The CME has pushed the streamer material aside as it escapes the corona. No bright leading edge is seen.}
              \label{fig:fig5}
\end{figure*}

The positions of STA and PSP on 27 January at 0 UT can be seen in Fig.~\ref{fig:fig0}. The first indication of the final CME eruption was seen in STA/COR2 on 25 January 2020 at $\sim$~13 UT on STA’s west limb. For about 8-10 hrs prior to this, swelling and brightening of the streamer could be seen, indicating that this was a SBO-CME. Thus, the timing of the CME is consistent with the CME trajectories determined from the WISPR data, which place the final ejection no later than mid-day January 25. 
The COR2 images in Fig.~\ref{fig:fig5} show, on the left,  the streamer before the eruption on 25 January at 0:2:39~UT and, on the right,  the streamer-blowout CME on 26 January at 00:36~UT. In the image on the right, streamer material is evident both above and below the cavity region as the CME pushes the streamer open during its escape.  No bright front is seen, indicating a slow CME, consistent with the average velocity of 263 km/s for  the trajectories in Section 2.1 (Table 1). By 26 January  at $\sim$ 3-4 UT, the CME is near the edge of the COR2 FOV, just as the CME enters the WISPR-I FOV. Unfortunately, STA has a data gap on 26 January  from 4 UT until 20 UT, by which time the CME is in the WISPR-O FOV.

\begin{figure}
   \resizebox{\hsize}{!}{\includegraphics {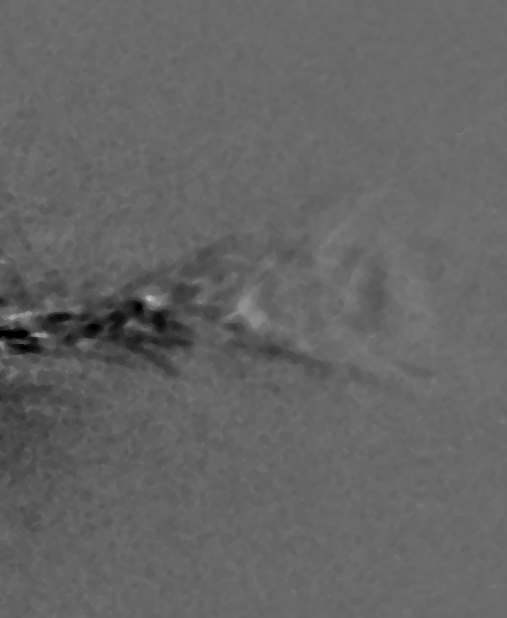}}
   \caption{A STEREO-A/ HI-1 cropped running-difference image of the CME on 26 January at 20:49~UT. This is close to the time of the WISPR image in Figure 2, lower panel and similar morphology is seen. An animation of the of CME as seen by HI-1 is available \href{http://sd-www.jhuapl.edu/secchi/paper_repo/movie2_20200126_27HI1Ardiffs.mp4}{online}.}
              \label{fig:fig6}
\end{figure}

Fig.~\ref{fig:fig6} shows the CME as seen in a  HI-1 cropped running-difference image on 26 January  at 20:49~UT. This is close to the time of the WISPR images in the lower panel of Fig.~\ref{fig:fig2}. The overall similarity of the structures is evident, e.g., a cavity region trailed by a large V-shaped structure, displaced streamer material, and no evidence of a bright leading edge or shock. A movie of the running-difference images covering 26-27 January is available \href{http://sd-www.jhuapl.edu/secchi/paper_repo/movie2_20200126_27HI1Ardiffs.mp4}{online}. The movie shows considerable outflow with blobs in the wake of the CME well after the CME leaves the FOV. This outflow was likely sampled in situ by PSP instruments a few days later when PSP encountered streamer plasma. There was no useful data from either COR1 or HI-2 for this event.

\begin{figure*}
   \centering
   \includegraphics[width = 18cm]{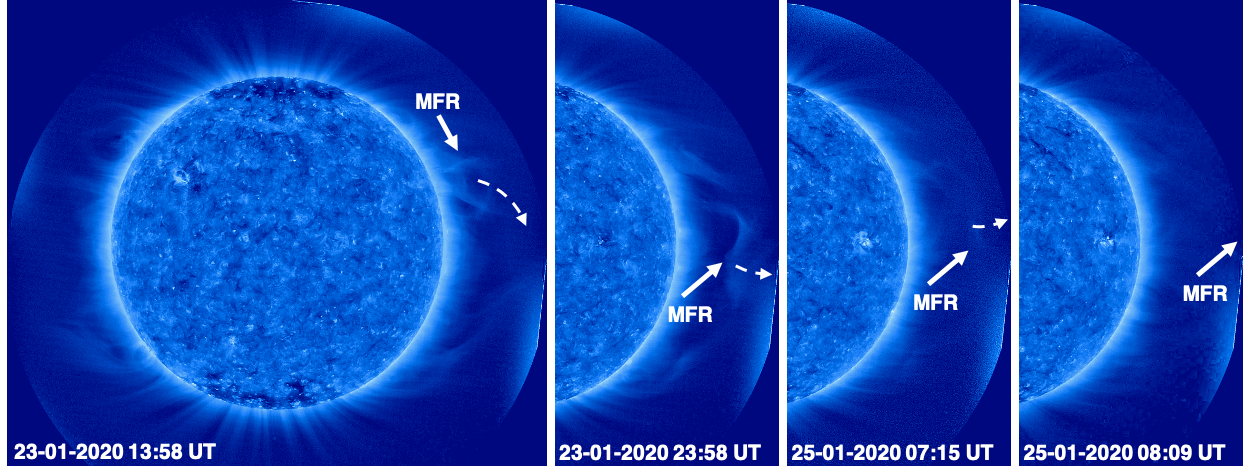}
   \caption{Snapshots of the low coronal evolution of the MFR associated with the 25 January, 2020 CME, as seen by STA/EUVI 171~\AA\ . The off-limb emission has been enhanced by removing the background stray light using a wavelet technique \citep{Stenborg2008}. The dashed arrows mark the direction of motion. The full movie is available \href{http://sd-www.jhuapl.edu/secchi/paper_repo/movie3_20200123_euvi_171wa_event.mp4}{online}.}
              \label{fig:fig7}
\end{figure*}

\section{Evolution of Source Region Leading to Flux Rope Formation, Rise,  and Eruption}
During this period, STA and SDO, in Earth orbit, were observing the Sun from vantage points separated by 79$^\circ$  (Fig.~\ref{fig:fig0}). They were positioned well to observe both the source region and the low corona evolution of the CME. In the following, we use off-limb observations from EUVI and on-disk observations by the Atmospheric Imaging Assembly \citep[AIA;][]{Lemen2012} and  Helioseismic and Magnetic Imager \citep[HMI; ][]{Schou2012} onboard SDO to identify the source region and probe the mechanisms leading to the eruption of the magnetic flux rope (MFR). As we will see in the following, our data analysis suggests a highly unusual eruption sequence.

\subsection{STEREO-A EUVI Observations }
The most likely candidate for the source of the 25 January CME seen in COR2 (and WISPR) is a magnetic flux rope structure that began to erupt a full two days earlier. On 23 January, an MFR-like structure rises from around 12:00~UT to at least 23:58~UT 
from a location in the northeast, as seen in EUVI 171\AA\ images (Fig.~\ref{fig:fig7} and \href{http://sd-www.jhuapl.edu/secchi/paper_repo/movie3_20200123_euvi_171wa_event.mp4}{online}). The structure moves towards the equator as it rises, in a ‘tumbling’ manner, common for slow eruptions during solar minimum \citep[e.g.][]{Panasenco2013}. Unfortunately, no SECCHI data is available from 23 January 23:58~UT through 24 January due to a data gap. The next available time series, on 25 January, from 00:07~UT on, shows the bottom part of a structure at the same location as the 23 January MFR (Fig.~\ref{fig:fig7} ). This structure can be seen to leave the EUVI field of view by about 6~UT on 25 January, several hours before the CME was detected in COR2. The  motions on 23 and 25 January, indicated by dotted arrows on Fig.~\ref{fig:fig7}, can be best seen in the \href{http://sd-www.jhuapl.edu/secchi/paper_repo/movie3_20200123_euvi_171wa_event.mp4}{online} movie.  An inspection of the online movie suggests that the 23 and 25 January structures are likely the same, but the lack of continuous observation due to the 24 January data gap, makes this interpretation somewhat uncertain. We did an exhaustive search of the available imaging data from LASCO and WISPR but found no evidence of eruptions between 23 January and our CME on the 25th. STA/COR2 and HI-1 suffer from the same data gap as EUVI. However, there should be some evidence in HI-1 of any events prior to our target CME due to the large HI-1 FOV (15 to 96 R$_{\odot}$, approximately). We found none. So we infer that the flux rope that started erupting slowly on 23 January at 11~UT in EUVI 171\AA \ was constrained by the overlying coronal fields until 25 January, probably rising slowly throughout. The first clear signature of the CME (the opening up of the steamer)  is seen at  $\sim$ 13~UT in the COR2 FOV. Such a slow evolution in the inner corona is not as unusual as it seems. \citet{Hess2020} reported a similarly slow rising MFR structure for the first CME captured by WISPR on 1 November 2018. These events are usually associated with streamer-blowout CMEs (SBO-CMEs), named after the almost complete evacuation of the streamer during the ejection. In an extensive analysis of these events, \citet{Vourlidas2018} found that the average SBO-CME duration is 40.5 hours. Therefore, our CME is fully consistent with a typical SBO-CME.

Slow rising CMEs leave very little trace of activity on disk. An SBO-CME was the archetypical ‘stealth’ CME \citep{Robbrecht2009}. The EUVI images show no connection to AR 12757, clearly seen on the disk by EUVI on 25 January, which suggests that the event originates behind the visible (from STA) western limb. We, hence, turn to the AIA and HMI observations to search for the source region.

 \begin{figure*}
   \centering
   \includegraphics[width = 17 cm]{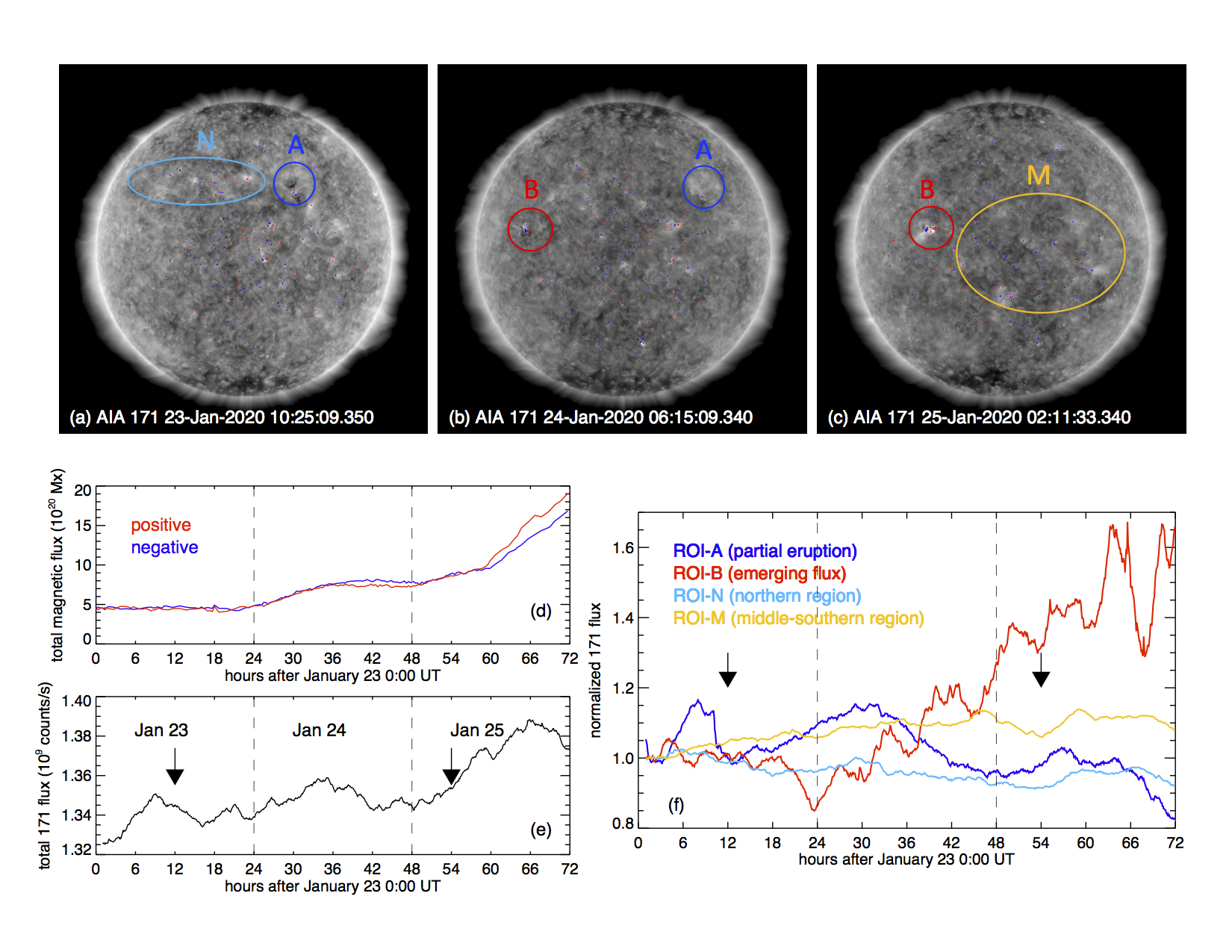}
   \caption{(a)-(c): Full disk SDO/AIA images in 171~\AA\ passband on 23, 24 and 25 January, overlaid with contours ($ \pm 200$ G, red and blue) of line-of-sight magnetic flux density obtained by HMI at around the same times of the AIA images.  (d): signed total magnetic flux integrated over the full disk. (e): total EUV flux in 171~\AA\ integrated over the full disk. (f): EUV flux in 171~\AA\ integrated in several Regions Of Interest (ROIs), as indicated by circles or ovals in panels (a) - (c). Note that the EUV flux is derived by integrating data counts in each of these ROIs with a fixed angular extent, and the changing projected area in the plane of sky due to solar rotation is not corrected. For clarity of display, these light curves are normalized to the flux at 0~UT on 23 January. Vertical dashed lines in panels (d-f) indicate the three stages of the radiation and magnetic activities, and the solid arrows indicate the times when the MFR was seen rising on 23 January and then erupted on 25 January, as observed in the STA images.An animation of the AIA/HMI overlays is available \href{http://sd-www.jhuapl.edu/secchi/paper_repo/movie4_ia171hmi_3day_halfdisk.mp4}{online}.}
              \label{fig:fulldisk}%
    \end{figure*}

\subsection{HMI and AIA Observations} 

We have analyzed full-disk EUV 171~\AA\ images obtained by SDO/AIA, and full-disk, line-of-sight magnetograms obtained by SDO/HMI from 23 January through 25 January. These observations reveal surface activities that are likely related to the evolution of the corona and flux rope during its eruption. HMI shows emergence and cancellation of magnetic flux, including the birth of AR 12757 prior to the CME. Prominent radiation signatures in the AIA 171~\AA\ passband are brightenings,  indicative of reconnection heating, and dimmings which indicate evacuation of coronal material due to expansion of the coronal structure  \citep{Sterling1997, Thompson2000, Reinard2009} or by interchange reconnection between low-lying and high-lying magnetic fields \citep[e.g.][]{Attrill2006, Downs2016}. Such signatures were observed at the times when the flux rope was seen by STA/EUVI to rise  on the morning of 23 January and when the flux rope escaped two days later on 25 January (Section 2.2). These signatures are likely
related to changes in the corona during the partial eruption and final escape of the flux rope. 

The top panel of Fig.~\ref{fig:fulldisk} shows a series of 171~\AA\ full-disk images during the three days, superimposed with the contours of the longitudinal magnetic field at $\pm 200$~G in red and blue, respectively. Also shown in Fig.~\ref{fig:fulldisk}d and ~\ref{fig:fulldisk}e are the signed total magnetic flux (in units of Mx) and the total 171~\AA\ flux (in units of counts per second), both integrated over the entire disk, over the three days. Brightenings and dimmings were prevalent in various regions within $ \pm$30$^{\circ}$ in latitude from the equator. We identify several of these regions as Regions of Interest (ROIs), which are denoted by color-coded circles or ovals in Fig.~\ref{fig:fulldisk}a-c. The 171~\AA\ light curves for these ROIs are given in Fig.~\ref{fig:fulldisk}f with the same  colors. The \href{http://sd-www.jhuapl.edu/secchi/paper_repo/movie4_ia171hmi_3day_halfdisk.mp4}{movie} shows more clearly radiation and magnetic activities in these regions during the three days, which may be described in three stages, as marked by the vertical dashed lines in Fig.~\ref{fig:fulldisk}d-f. We discuss each stage in detail below.

{\bf Formation and Rise of the Flux Rope: } 
Throughout 23 January, the total magnetic flux does not vary significantly, but the total 171~\AA\ flux increases from the start of the day until 10~UT, and then decreases. Specifically, sudden brightenings with small ejecta occurred before 3~UT on 23 January in a bipolar region (referred to as Region of Interest-A, or ROI-A hereafter) in the northwest, 
shown in Fig.~\ref{fig:fulldisk}a. Subsequently, from 8~UT, significant dimming was observed in ROI-A. From 6 to 12 UT, a couple of bipolar regions to the south-east of ROI-A, near the equator and central meridian, also exhibit brightenings. Finally, a sequence of brightenings and dimmings, although not as prominent, was observed in extended areas across the disk along the similar latitude, or the northern region ROI-N, shown in Fig.~\ref{fig:fulldisk}a. The lightcurves in 171~\AA\ in ROI-A and ROI-N are given in Fig.~\ref{fig:fulldisk}f. This series of events coincided with the appearance of the rising flux rope in the field of view of EUVI (Fig.~\ref{fig:fig7}), and are therefore likely the on-disk manifestations of the partial eruption of the MFR see on the limb by EUVI.

{\bf Emergence of AR 12757: }
Early on 24 January, emerging flux was observed 
near the east limb as seen from Earth (and near the west limb as seen from STA). The flux emergence region is referred to as ROI-B and denoted in Fig.~\ref{fig:fulldisk}b and c. In twelve hours, the magnetic flux in ROI-B has increased by 4$\times 10^{20}$ Mx in both the positive and negative polarities (Fig.~\ref{fig:fulldisk}d), associated with prominent alternating brightenings and dimmings in this area as shown in Fig.~\ref{fig:fulldisk}f. The flux emergence paused at 12~UT on 24 January for the remainder of the day, although brightenings and dimmings continue to occur. This region was officially identified as AR 12757 on 25 January.

{\bf Ejection of the Flux Rope:}
Early on 25 January, rapid flux emergence took place again in AR 12757 (ROI-B), as shown in Fig.~\ref{fig:fulldisk}d. This was accompanied by brightenings due to formation and heating of new loops in the emerging active region. Subsequently, dimmings occurred around these new loops and are visible in the included \href{http://sd-www.jhuapl.edu/secchi/paper_repo/movie4_ia171hmi_3day_halfdisk.mp4}{movie}.  These signatures suggest reconnection between emerging flux and adjacent or overlying fields, leading to a restructuring of the corona \citep[e.g.,][]{Shibata1992, MorenoInsertis2013}. Weak coronal dimmings were also observed in an extended area, denoted as the middle-southern region, ROI-M,  in Fig.~\ref{fig:fulldisk}c and f, which proceeded from the start of the day till 6~UT on 25 January. These events coincided with the final eruption of the flux rope, seen leaving the EUVI field of view by 6 UT on 25 January near the solar equator (Fig.~\ref{fig:fig7}, and accompanying movie). These signatures are likely related to the final eruption of the CME by restructuring global coronal magnetic fields constraining the CME. On the other hand, they are not direct signatures of an ejection from the low corona, which should usually produce significant and expanding dimmings, waves, or cool ejecta, not evident in the disk observations.  This is consistent with the release and acceleration of the CME high in the corona and lends further support to our eruption scenario. The lack of association between AR 12757 and the rising coronal structure can also be seen in the EUVI \href{http://sd-www.jhuapl.edu/secchi/paper_repo/movie3_20200123_euvi_171wa_event.mp4}{movie}  on 25 January; AR 12757 is the small bright region approaching the west limb.

 \begin{figure*}
   \centering
   \includegraphics[width = 15cm]{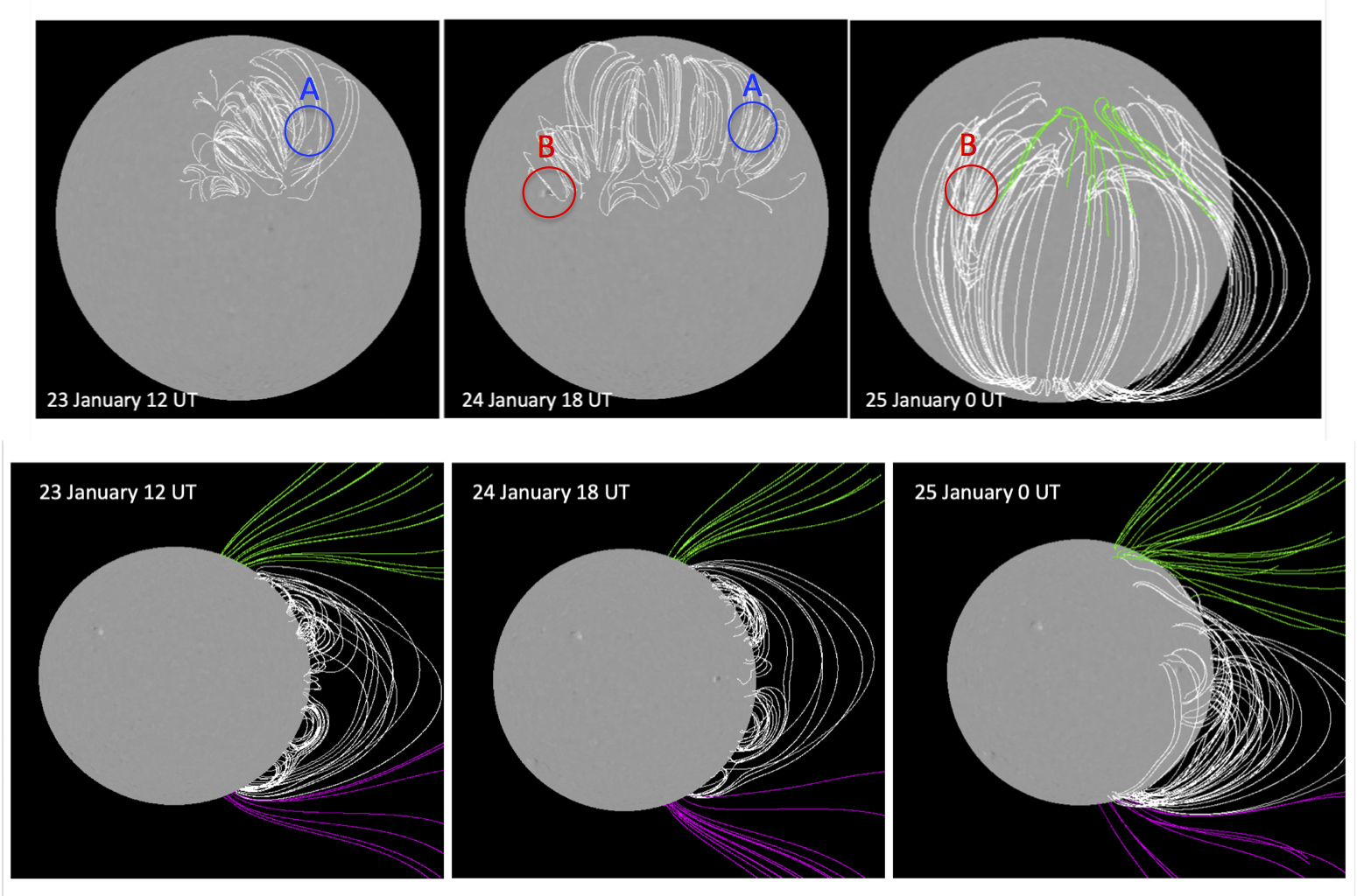}
   \caption{PFSS reconstructions of the coronal magnetic fields at three times showing the evolution of the global magnetic field from the SDO view (top panels) and from the STA view (lower panels). The three times, 23 January at 12 UT, 24 January at 18 UT, and 25 January at 0 UT, show the global magnetic field during the three stages described in the previous text. The green and magnetic lines show the open field lines of two polarities. }
              \label{fig:pfss}%
    \end{figure*}

\subsection{Comparison with PFSS modeling}

The combined STEREO and SDO observations provide a comprehensive view of CME-associated activities, which occur over a large range of heliolongitudes. 
To probe the connection between the evolution of the flux rope and brightening and dimming signatures on the disk, we reconstruct the global magnetic field using the PFSS model available in SolarSoft. This  model uses synoptic magnetograms which include the effects of magnetic flux transport as described in \citep{Schrijver2003},  updated to use HMI magnetograms. New active regions are not incorporated into the synoptic magnetograms until reaching the HMI FOV which covers about $\pm 30^{\circ}$ from the central meridian. Thus AR 12757, which emerged on 24 January, is not in the synoptic magnetograms until 25 January .

Fig.~\ref{fig:pfss} 
shows the global magnetic field from the SDO view (upper row) and from the STA view (lower row), during the three stages described in the previous section. Before the emergence of AR 12757, the maps reveal a magnetic arcade extending from the north-west toward the center of the Sun with multiple smaller arcades underneath. These arcades indicate polarity inversion lines over which  flux ropes could form. After 24 January, the arcade is further elongated toward the east. The ROI-A exhibiting prominent brightenings and dimmings at the time of the partial eruption of the flux rope on 23 January is located at the western end of the arcade, whereas the emerging AR 12757 (ROI-B) is located at the eastern end of the elongated arcade.

These observations suggest that the flux rope was likely confined in the corona, and its partial eruption on 23 January and final ejection on 25 January were related to activities in these two ROIs. It is plausible that, during the initial eruption on 23 January, the rise of the flux rope was held back by overlying coronal magnetic fields.
The brightening and dimming signatures in the northern band (ROI-N) probably reflect the impact of the rising flux rope on the overlying coronal fields. Two days later, interaction between the emerging active region and magnetic field of the flux rope or the large scale coronal fields overlying the flux rope probably played a role in the final MFR ejection. 

Interestingly, the PFSS reconstruction on 25 January, one day after the emergence of AR 12757, shows significant changes in the global configuration with a large amount of magnetic flux opening up in the northern hemisphere (Fig.~\ref{fig:pfss}, top panel, right). This change also agrees with the hypothesis that the flux rope was lying somewhere along the east-west direction north of the equator, and its eruption opened up the magnetic field there. 

To summarize, although the newly emerging AR 12757 has probably played an important role in the final ejection of the CME on 25 January, joint STA and SDO observations reveal that the flux rope has started to rise two days earlier, before the birth of AR 12757. Based on the tracking and fitting trajectories in Table 1, the CME is displaced by at least 30$^{\circ}$ to the west of AR 12757. The trajectories were determined from the WISPR data using a tracking and fitting technique which assumes a constant radial motion. To further examine the location of the CME, we also apply triangulation using CME observations from multiple views.

 \begin{figure*}
   \centering
   \includegraphics[width = 18cm]{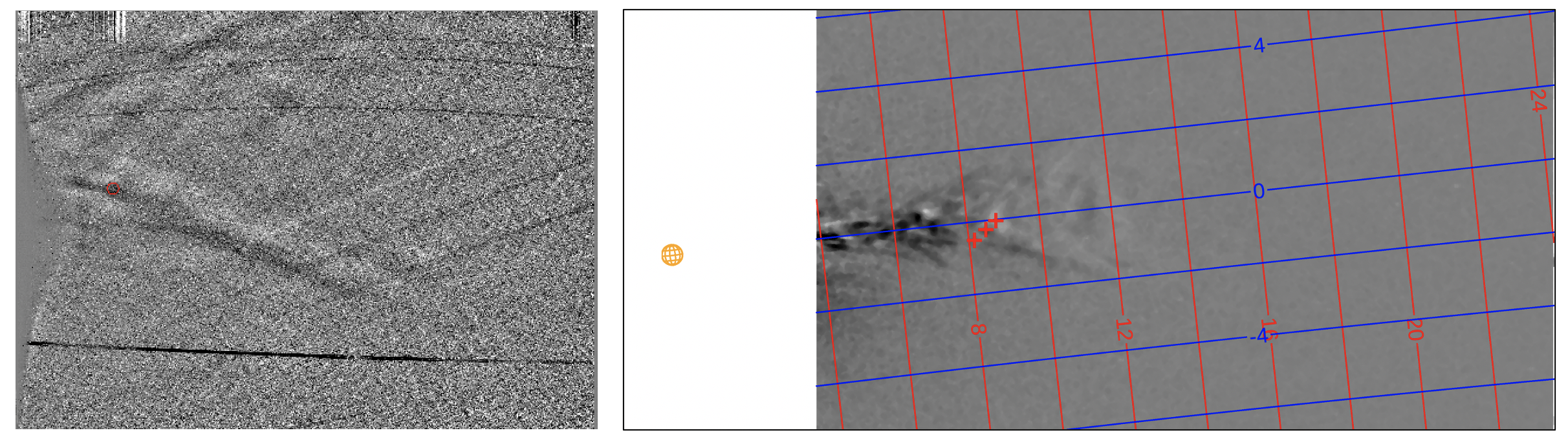}
   \caption{Left: WISPR-O image of 26 January at 20:49 UT with the selected feature circled in red. Selecting this feature defines a line-of-sight from PSP to the feature. Right: Simultaneous HI-1 image with points along the line sight from PSP to the selected feature for three distances along this line-of-sight, projected onto the image (red +’s). The points correspond to distances of 10, 27 and 44 R$_{\odot}$; the point corresponding to 27 R$_{\odot}$ falls on the feature selected in the WISPR-O image, which determines the feature's coordinates (see text). The Sun (yellow globe) in the panel on the right is shown to scale. 
   The HI-1 image is projected in the Helioprojective Cartesian (HPC) system (red and blue grid lines) with the Sun (yellow globe) drawn to scale.}
              \label{fig:triangle}
\end{figure*}

\section{Determining CME’s Position by Triangulation using WISPR and STEREO-A}
 
When an object or  feature is observed simultaneously from two spacecraft, it may be possible to use triangulation to determine the 3D location of the feature at that time. Unlike the tracking and fitting technique used to determine the CME trajectory (position vs. time) presented in Section 2, triangulation makes no assumption about the motion of the feature.  It does assume that the same “feature” can be seen from both viewpoints, which works best for localized features. Usually, because of the diffuse nature of CMEs and the line-of-sight integration of coronagraphs, the CME may look quite different from the two viewpoints and the application of this technique to coronagraph images is limited \citep{Liewer2009, Colaninno2013}. However, in this event, both WISPR and STA images resolve structure within the CME and thus several features can be identified in simultaneous image pairs. Because of the STA data gap on 26 January from 4  to 20  UT,  simultaneous images exist only for a few hours.

 We use a novel triangulation technique to determine the location of the CME using simultaneous FITS images from STA and WISPR that relies on the SolarSoft package \href{https://sohowww.nascom.nasa.gov/solarsoft/gen/doc/aspice/sunspice.pd}{SUNSPICE}. This technique can be used to determine the 3D location of any feature in a WISPR image that is seen simultaneously in an image from a white-light telescope on another spacecraft provided the FITS images have both proper World Coordinate System   \citep[WCS;][]{Thompson2006} and camera model information in their headers. To start, the user selects the feature in the WISPR image. The pixel location in the image defines a particular line-of-sight through space; the WCS and camera model information in the header allow that line-of-sight to be traced in, e.g., the Heliospheric Inertial frame.  The feature could lie anywhere along that line-of-sight. Any point along that line is specified by its distance $d$ from PSP. Next, an array of possible distances of the feature from PSP is chosen, which then specifies an array of points along the line-of-sight, here,  in HCI coordinates. These points are then projected onto the simultaneous image from the second spacecraft using WCS and camera model information in its FITS header. When one of these points projects onto the same feature in the second image, the HCI coordinates of this feature have been found. Trial and error are used to find the proper distance of the feature from the spacecraft. If none of the projected points falls on the feature in the second image, the process is repeated with a new array of distances until one of the projected points “hits" the feature in the second image. Once the proper distance to the feature has been determined, the HCI coordinates of this point represent the 3D location of the feature.
 
This technique is illustrated in Fig.~\ref{fig:triangle}, using images from STA/HI-1 and WISPR-O. The left image is the WISPR-O FITS image on 26 January at 20:49 UT. We use the same WISPR-O feature used in the trajectory determination: the dark spot behind the diffuse V. The  position in the image of the selected feature (red circle) specifies WISPR-O’s line of sight to the feature. The right image is the HI-1 image at the same time, showing the projected locations of three points along the WISPR line-of-sight, as viewed from STA, for three assumed distances $d$ from PSP. The projected locations (red +’s) were specified by the three distances from PSP of $d$ = 10, 27 and 44 R$_{\odot}$.  The middle red +, corresponding to the distance of $d$=27 R$_{\odot}$, falls on the selected feature, and, thus the distance to this feature from PSP is determined. The uncertainty in this distance is 2 R$_{\odot}$, found by varying the distance as much as possible while the red + still appeared to “hit” the feature.  The coordinates of the feature found in this manner are distance from the Sun r/R$_{\odot}$ = 31 $ \pm $ 2 , HCI longitude and latitude = (  $66^{\circ} \pm  3^{\circ}$, 
$-2^{\circ}  \pm 2^{\circ}$). These angles are in excellent agreement with the angles found by tracking and fitting this feature in Table 1 (average angles:  HCI longitude and latitude = ($61^{\circ} \pm 5^{\circ}, 1^{\circ} \pm  2^{\circ}$). The distance from the Sun is also in excellent agreement with the predicted distance at this time of r/R$_{\odot}$= $31.2 \pm   0.3$, validating our trajectory solution.

 \begin{figure*}
   \includegraphics[width = 18 cm]{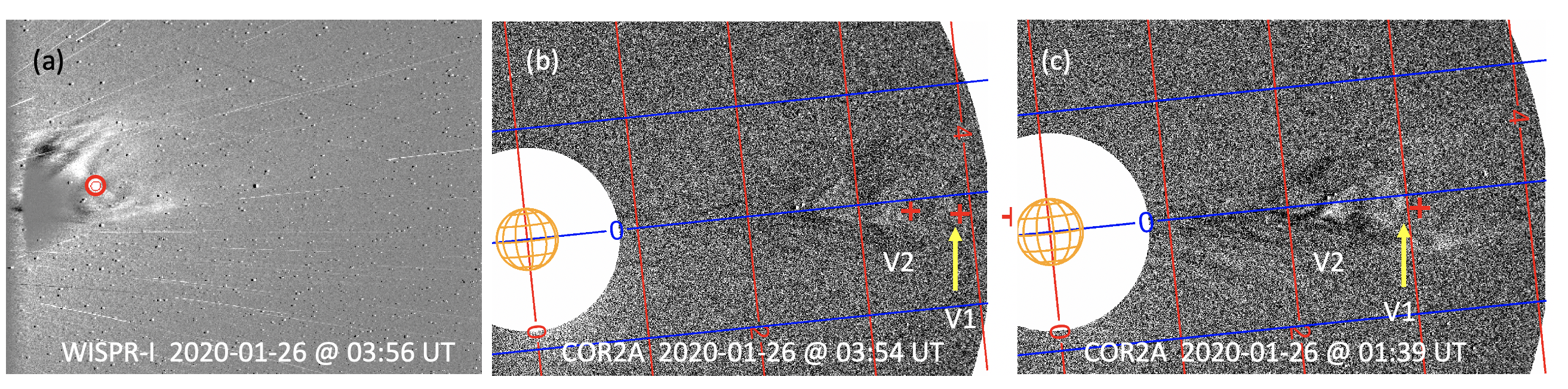}
   \caption{Left: WISPR-I running differenced image at 03:56 UT with the feature  used in tracking (bright spot at the back of the cavity) circled in red. Center: nearly simultaneous STA/COR2 running differenced image (03:54  UT) with the common feature use for triangulation labeled v1/yellow arrow.  Also shown are three predicted locations (red +’s) of the WISPR feature projected onto the COR2 image for three possible distances of the feature from PSP. The middle red + falls on the feature and thus determines the 3D location of the common feature (see text). Right: COR2 image at an earlier time (01:39 UT) when the full CME cavity structure was visible, allowing identification of the feature (marked v1/yellow arrow).
   The COR2 image is projected in the Helioprojective Cartesian (HPC) system (red and blue grid lines) with the Sun (yellow globe) drawn to scale.
 }
              \label{fig:cor2triangle}
\end{figure*}
We also used the above triangulation technique to determine the location of the CME at an earlier time (26 January at 03:54 UT) using WISPR-I and STA/COR2 images. The only simultaneous WISPR-COR2 image pairs occur when the CME was just entering the WISRP-I FOV and just leaving the COR2 FOV. The two simultaneous images used in the triangulation are shown in Fig.~\ref{fig:cor2triangle}a (WISPR-I at 03:56 UT) and Fig.~\ref{fig:cor2triangle}b  (COR2 at 03:54). Both are running difference images. The CME feature used in the triangulation is the same bright V at the back of the cavity (Fig.~\ref{fig:fig3}) used for  trajectory determinations (Table 1); it is circled in red in Fig.~\ref{fig:cor2triangle}a. At this time, the full structure of the cavity was no longer visible in the COR2 image (Fig.~\ref{fig:cor2triangle}b), but several bright V-shaped features are visible. It was only possible to determine which bright V corresponded to the circled bright V feature in the WISPR-I image by using the sequence of COR2 images leading up to this time. We started at a time more than two hours earlier (01:39 UT) when the cavity structure was still visible (Fig.~\ref{fig:cor2triangle}c). In this image it is clear which bright feature marks the back of the cavity:  Note the similarity in the relation of this feature, labeled V1/yellow arrow) to the CME cavity to that of the circled feature and cavity in the WISPR image at 03:56 UT (Fig.~\ref{fig:cor2triangle}a).  Once the common feature was identified at this earlier time, we followed it in the sequence of five COR2 images to the time of the simultaneous pair (03:54 UT, Fig.~\ref{fig:cor2triangle}b) where the feature (V1, yellow arrow) has now moved towards the edge of the COR2 FOV. In addition, we then verified that we had identified the correct feature in the COR2 image  by using the trajectory determined for this feature in Sec. 2.1 (Table 1) to predict  the HCI coordinates of this feature at the time of the earlier image, before it was visible in WISPR, and projecting this location on to the earlier COR2 image  at 01:39 UT. The predicted location of this feature is shown as the red + in  Fig.~\ref{fig:cor2triangle}c. It can be seen that it falls quite close to the bright feature (V1/yellow arrow). Considering that this image is at a time 2 hrs before the tracking in WISPR began, we consider our feature identification to be reliable.

With a common feature identified in the simultaneous image pair, we now apply the triangulation technique used for the HI-1 triangulation described above. We projected potential locations for the WISPR feature V1(Fig.~\ref{fig:cor2triangle}a, inside red circle) at 03:56 UT onto the nearly simultaneous COR2 image (03:54 UT) for an array of distances $d$ from PSP. Fig.~\ref{fig:cor2triangle}b shows the projected location for three assumed distances of the feature from PSP: $d$/R$_{\odot}$ = 28, 33, and 38. As for the HI-1 triangulation, the uncertainty was determined by varying the choice of distances over which the projected point appeared to fall on the feature. The end result for the location of this feature was distance from the Sun r/R$_{\odot}$ = $14.0 \pm  0.1$, HCI longitude and latitude = ($67^{\circ} \pm  5^{\circ}, -2^{\circ} \pm  1^{\circ}$). 
Recall that the trajectory determined for this feature in Sec. 2.1 had a smaller HCI longitude ($57^{\circ} \pm  2^{\circ}$). 
The trajectory predicts that this feature should be at r/R$_{\odot}$ = $14.5 \pm 0.1$  at this time.  The agreement between the triangulated location and the predicted location from the trajectory for this COR2 case is good, but not as good as in the HI-1 case and not within the error bars due to the techniques used. This may be due to the line-of-sight integration problem for coronagraph images discussed above.  Also, as discussed in \citet{Liewer2020}, there are additional sources of errors in the tracking and fitting technique itself, most obviously from the possible violation of the assumption of radial propagation at a constant velocity. We conclude that the two techniques, tracking and fitting versus triangulation, are complementary and help establish the overall limits on both techniques and help improve the usefulness of both.

 
\section{Summary and Discussion}
 
In this paper, we have presented a detailed analysis of the evolution of a streamer-blowout CME on 25 January 2020 using data from three spacecraft: PSP, STEREO-A and SDO. WISPR on PSP provided images of the propagation of the CME and its internal structure from a distance of about 30 R$_{\odot}$, revealing a complex flux rope morphology. The WISPR data were also used to determine the trajectory of the CME using the technique of \citet{Liewer2020}. STA/EUVI provided data that suggested that the magnetic flux rope partially erupted two days before the final ejection of the CME on January 25, 2020.  Unfortunately, there was a STA data gap on 24 January, so we are unable to follow the flux rope continuously. However, a thorough search of available data (STEREO, SOHO, PSP) did not show any eruptions on 23-24 January.  On 25 January, STA/COR2 data showed both the swelling of the streamer beforehand and the final ejection of the CME, which led us to classify the event as a streamer-blowout CME. STA provided images of the  CME’s propagation, including simultaneous images with WISPR at several times. These simultaneous images allowed us to use a novel triangulation technique to determine the CME’s  positions at the times of the image pairs, confirming the trajectory determination using the WISPR data alone. The technique can be used to locate features seen in simultaneous FITS images from WISPR and any second spacecraft, provided the images have the proper World Coordinate System \citep{Thompson2006} and camera model information in their headers.
 
The CME was near STA’s west limb and, fortuitously, SDO was about 80$^{\circ}$ to the west of STA and provided excellent coverage of the underlying solar surface activity.
SDO/HMI data was used to track magnetic flux emergence and disappearance for the three days 23-25 January, 2020. 
SDO/AIA 171~\AA\ data was used to track brightenings and dimmings, proxies for coronal heating by reconnection and the opening and/or rising of coronal loops (leading to evacuation or cooling), respectively.
Three stages of activity were analyzed: the confined eruption on 23 January, the emergence of new active region AR 12757 on 24-25 January, and the final ejection of the magnetic flux rope early on January 25. The SDO analysis was supplemented by PFSS modeling of the corona corresponding to each of the three stages.  

The analysis supports the following scenario: A magnetic flux rope partially erupted on 23 January.
The prominent brightening and dimming observed in a bipolar regions in the NW quadrant by SDO were possibly related to the initial eruption.
The flux rope deflected southward (cf. (Fig.~\ref{fig:fig7}), guided toward the heliospheric current sheet near the solar equator by the overlying magnetic fields \citep{Liewer2015}, which also confined the eruption (Figure~\ref{fig:pfss}). On 24 January, a new active region, later NOAA  AR 12757, started emerging in the NE quadrant as seen from SDO. This led to a major restructuring of the solar magnetic field according to the PFSS models. Apparently, this reduced the constraints on the flux rope, leading to the brightening and swelling of the overlying streamer and final eruption of the flux rope.  This scenario is supported by the trajectory determined from the WISPR data using the tracking and fitting technique of \citet{Liewer2020}: The CME’s HCI latitude was found to be $2^\circ \pm  2^\circ$ and is consistent with the latitude of the very flat heliospheric current sheet essentially at the solar equator during this time period. The longitude of about $61^\circ \pm  4^\circ$ puts the trajectory about $13^\circ$ west of Earth and about $35^\circ$ west of the location of AR 12757 on 25 January. 
The HCI latitude of $1^\circ \pm 2^\circ$ converts to an HEE latitude of $8^\circ \pm 2^\circ$. The CME could have given Earth a glancing blow, but no ICME signatures were observed in situ (Q. Hu, private communication). The average velocity of the CME was found to be v=263$ \pm $20 km/s.  \citet{Nindos2020} (this issue) also made a determination of the velocity of this CME as part of their work on tracking WISPR-I features. They found v = 258 km/s (their Fig. 6, track Number 4), in excellent agreement with our result.

A prolonged initiation phase for slow CMEs is not uncommon. As noted above, \citet{Vourlidas2018} derived an average duration for SBOs of 40.5 hours. Many polar crown filaments exhibit the characteristic shape of a flux rope and can exist for days before erupting \citep{Gibson_2015}. For example \citet{howard_formation_2014} analyzed such a case and found that the slow (~400km/s), CME launched about 5 days after the formation of the flux rope.
 
We note that the \href{https://kauai.ccmc.gsfc.nasa.gov/DONKI}{DONKI} data base also gave a prediction for the 25 January 2020 CME to deliver  Earth a  glancing blow. The CME speed in the database is higher (v=362 km/s), the start time later, and the source region is identified as the active region AR 12757. As we discussed above, this AR was in the FOVs of both SDO and STA and analysis of AIA and EUVI data showed little connection between the CME and this AR. The most likely reason for the misidentification is the extremely slow evolution of this CME. It is not customary to search for CME eruptions days before a CME while it is customary (and human nature) to tend to associate CMEs with the nearest on-disk activity. As we show here \citep[and in the past, e.g.][]{Nieves-Chinchilla2013, 
Patsourakos2013, Colaninno2015}, 
the easiest approach is not necessarily the correct one. This is particularly true for solar minimum CMEs that tend to have weak or even no obvious low surface manifestations. Our event falls under the ‘stealth’ CME category. As for the original ‘stealth’ CME \citep{Robbrecht2009}, the event itself and its source region were identified only thanks to the off-”Sun-Earth” Line (SEL) EUVI and coronagraphic observations from STA. The 26 January CME event demonstrates two important space weather aspects of slow CMEs: 
the formation/initial eruption phases of such events can be long-lived, lasting for more than a day, below the traditional FOVs of coronagraphs. Care must be exercised in the identification of the source regions with the search extending at least a day or two prior to the identification of the eruption in the middle corona.
Although slow CMEs are less likely to be geo-effective on their own, they carry magnetic fields and will interact with other CMEs or stream interface regions (SIR), altering their magnetic and plasma configuration. A slow CME, for example, could be trapped and compressed at the leading edge of an SIR \citep{rouillard2011}, increasing the potential impact on geospace. So it is important to include such events in the operational modeling of the inner heliosphere. The most efficient way to detect them are the off-SEL observations, say, from Lagrange L4 or L5 vantage points. 


\bibliographystyle{aa}
\bibliography{E4jan26.bib}

\begin{acknowledgements}
We gratefully acknowledge the help and support of the WISPR team and William Thompson, Adnet System, Inc., throughout this work.
The work of P.~C.\,Liewer, J.~R.\,Hall, and P.\,Penteado was conducted at the 
Jet Propulsion Laboratory, California Institute of Technology under a contract from NASA. 
A.\,Vourlidas is supported by the WISPR Phase-E funding and NASA grant 80NSSC19K1261. 
J.\,Qiu is partly supported by NASA's HGI program (80NSSC18K0622). 
Parker Solar Probe was designed, built, and is now operated by the Johns Hopkins Applied Physics Laboratory as part of NASA’s Living with a Star (LWS) program (contract NNN06AA01C). Support from the LWS management and technical team has played a critical role in the success of the Parker Solar Probe mission.
\end{acknowledgements}

\end{document}